\documentclass[usenatbib,useAMS,preprint,psfig2]{mn2e}
\usepackage{times}
\usepackage{amsmath}
\usepackage{amssymb,amsfonts}
\usepackage{graphicx}
\usepackage{epstopdf}
\usepackage{verbatim}
\input psfig.sty

\def\gapprox{\lower.4ex\hbox{$\;\buildrel >\over{\scriptstyle\sim}\;$}}
\def\lapprox{\lower.4ex\hbox{$\;\buildrel <\over{\scriptstyle\sim}\;$}} \def\be{\begin{equation}}
\def\be{\begin{equation}}
\def\ee{\end{equation}}
\def\bea{\begin{eqnarray}}
\def\eea{\end{eqnarray}}
\catcode`\@=11
\font\tenmib=cmmib10
\font\tensyb=cmbsy10
\font\tenbi=cmmib10 
\newfam\bifam  
\textfont\bifam=\tenbi  

\def\unboldmath{\everymath{}\everydisplay{}
          \textfont\@ne\teni 
          \textfont\tw@\tensy
          }
\def\boldmath{$\!\!$\relax\everymath{\mit}\everydisplay{\mit}
        \textfont\@ne\tenmib
        \textfont\tw@\tensyb 
        \relax}%
\catcode`\@=12

\begin{document}
\title[From intensity to E-field]{Electric field representation of pulsar intensity spectra}
\author[Walker \&\ Stinebring]{M.A.~Walker$^{1,2,3}$, D.R.~Stinebring$^4$\\
1. School of Physics, University of Sydney, NSW 2006, Australia\\
2. Kapteyn Astronomical Institute, University of Groningen, P.O. Box 800, 9700 AV Groningen,
The Netherlands\\
3. Netherlands Foundation for Research in Astronomy, P.O. Box 2, 7990 AA Dwingeloo, The Netherlands\\
4. Oberlin College, Department of Physics and Astronomy, Oberlin, OH 44074, U.S.A.}

\date{\today}

\maketitle

\begin{abstract}
Pulsar dynamic spectra exhibit high visibility fringes arising from
interference between scattered radio waves. These fringes may be
random or highly ordered patterns, depending on the nature of the
scattering or refraction. Here we consider the
possibility of decomposing pulsar dynamic spectra -- which are intensity
measurements -- into their constituent scattered waves, i.e. electric field
components. We describe an iterative method of achieving this
decomposition and show how the algorithm performs on data from
the pulsar B0834+06. The match between model and observations
is good, although not formally acceptable as a representation of the
data. Scattered wave components derived in this way are immediately
useful for qualitative insights into the scattering geometry. With some
further development this approach can be put to a variety of uses,
including: imaging the scattering and refracting structures in the interstellar
medium; interstellar interferometric imaging of pulsars at very high angular
resolution; and mitigating pulse arrival time fluctuations due to interstellar
scattering.
\end{abstract}

\begin{keywords}
pulsars: general --- ISM: general --- scattering --- turbulence --- techniques: interferometric
\end{keywords}

\section{Introduction}\label{sec:intro}

In the last few years there has been much progress made
in understanding the interference fringes which are manifest
in pulsar dynamic spectra. Representing the data as power
spectra (``secondary spectra'') of the dynamic spectra
demonstrated the underlying simplicity of the
complex organised fringe patterns which are sometimes seen:
commonly it is found that the power is concentrated on
parabolic loci in the transform domain (Stinebring et al 2001).
This property is now understood in the following terms
(Stinebring et al 2001; Walker et al 2004; Cordes et al 2005).
The dynamic spectrum
is the electric field intensity, $I(\nu, t)$, as a function of
radio-frequency, $\nu$, and time, $t$; the conjugate
(Fourier) variables are the delay, $\tau$, and Doppler-shift,
$\omega$, of the scattered waves. The purely geometric
component of the delay is proportional to the square of the
scattering angle, whereas the Doppler-shift is proportional
to one component (the component parallel to the effective
transverse velocity vector) of the scattering angle, leading
to parabolic relationships between $\tau$ and $\omega$.

The well-defined parabolae which are often seen further
require that the scattering material not be distributed along
the line-of-sight, but instead must be concentrated into a
thin ``screen'' (Stinebring et al 2001). In some cases the
data also suggest that the scattering is highly anisotropic
(Walker et al 2004; Cordes et al 2005).

To date the interpretation of observed dynamic/secondary
spectra has proceeded by forward theoretical modelling;
in other words models are constructed and the predicted
secondary spectra are compared with the data. This approach
has yielded important insights, but it is fundamentally limited
in that a detailed match to the data is not practicable --- only
the general characteristics can be considered. This is a severe
limitation because dynamic spectra may contain a great wealth
of detailed information in the $\sim10^6$ independent
frequency-time measurements which can be routinely recorded.
Consequently we are motivated to attempt inverse modelling
of dynamic spectra, whereby the electric fields are deduced
by modelling the observed intensity distribution. If the total
electric field is $U(\nu, t)$, then $I(\nu, t)=U^*(\nu, t)\,U(\nu, t)$,
so $|U(\nu, t)|=\sqrt{I(\nu, t)}$ tells us the field amplitudes,
but the phases remain a priori unknown. This is an example
of a ``phase retrieval problem''; such problems are common
in the optics literature (e.g. Fienup 1982; Elser 2003; McBride,
O'Leary and Allen 2004), although we were unaware of that
resource until the present work was largely complete.

This paper presents an iterative approach to solving the
inversion problem; it is not the only approach, nor is it necessarily
the best approach, but it has been at least partially successful.
The outline of the paper is as follows: in the next section we
describe the physical model we have adopted and derive the
mathematical basis for our inversion algorithm; our attempt
at implementing this algorithm is described in \S3, while in
\S4 we show how this implementation performs on real data;
possible applications of these techniques are discussed in \S5.

\section{Foundation of the inversion}
As radio-waves propagate through the Galaxy they are
subject to refraction and scattering by inhomogeneities
in the ionised interstellar medium (e.g. Rickett 1991;
Narayan 1992). If the radio-waves originate
from a pulsar then the size of the source is small, and the
coherent patch correspondingly large, so that essentially all
pairs of scattered/refracted waves yield high-visibility interference
fringes, regardless of the spatial separation of the scattering
centres. Consequently the combination of scattered waves can
be well approximated simply by addition of the various electric
field components. (We consider departures from this point-like
source approximation in \S5.3.) Denoting the total
electric field as $U(\nu, t)$, and each of the discrete
scattered waves as $u_j(\nu, t)$, we have
\begin{equation}
U(\nu, t) = \sum_j u_j(\nu, t) = \sum_j \widetilde{u}_j\exp\left[ 2\pi i (\nu\tau_j + \omega_jt) \right].
\end{equation}
Equivalently, we can write this relationship in the Fourier Transform
domain, $(\nu, t)\rightarrow(\tau, \omega)$, as
\begin{equation}
\widetilde{U}(\tau, \omega) = \sum_j\widetilde{u}_j\,\delta(\tau-\tau_j)\,\delta(\omega-\omega_j),
\end{equation}
where $\delta$ denotes the Dirac delta-function, and the
$\widetilde{u}_j$ are complex constants. The intensity observed at radio
frequency $\nu$ (relative to band centre) and time $t$ (relative
to the central epoch of the observation) is $I(\nu, t) = U^*(\nu, t)\,U(\nu, t)$.
Given an observed $I(\nu, t)$, we want to determine $U(\nu, t)$.
More precisely, because the conjugate variables $(\tau, \omega)$
have clear physical interpretations, as the delay and Doppler-shift
of the scattered waves, we wish to determine $\widetilde{U}(\tau, \omega)$.
In this model the individual scattered waves which make up $\widetilde{U}$
are completely specified by their delay ($\tau_j$), Doppler-shift
($\omega_j$), amplitude and phase (the single complex number $\widetilde{u}_j$).

We have taken an iterative approach to solving for $\widetilde{U}$,
given $I$, as we now describe.
Given a model electric field, $U_0$, we can compute the
corresponding intensity pattern. In general this will not match the
data exactly and we want to improve the model. Suppose
our model differs from the true electric field by an amount $\delta U_0$,
we can then write
\begin{equation}
I(\nu, t) = U^*U = U_0^*U_0 + \delta U_0^* U_0 + U_0^* \delta U_0 + \delta U_0^* \delta U_0.
\end{equation}
(Where there is no ambiguity we will henceforth not make explicit the
independent variables.) Introducing the residual between model
and data, ${\cal R}_0(\nu, t) := I - U_0^*U_0$, we arrive at
\begin{equation}
{\cal R}_0 = \delta U_0^* U_0 + U_0^* \delta U_0+\delta U_0^* \delta U_0.
\end{equation}
Providing the existing model is a good one $\left(|\delta U_0|\ll|U_0|\right)$ we can
neglect the last term in this expression and estimate the
quantity $\delta U_0$ from the resulting linear approximation.

Now suppose that $\delta U_0$ is dominated by a single scattered wave,
so that our task reduces to a determination of the properties of that wave,
then we can make the approximation
\begin{equation}
\delta U_0(\nu, t) = \delta\widetilde{U}_0\exp\left[ 2\pi i (\nu\tau + \omega t) \right],
\end{equation}
for some particular (but unknown) values of $(\tau, \omega)$, where
$\delta\widetilde{U}_0$ is a complex constant. Dropping the final term
in equation 4, multiplying by $U_0(\nu, t) \exp\left[ -2 \pi i (\nu\tau + \omega t ) \right]$
and integrating yields
\begin{eqnarray}
\int\!\!{\rm d}t\,{\rm d}\nu\,{\cal R}_0 U_0 \exp\left[ -2 \pi i ( \nu\tau + \omega t ) \right] =
\qquad\qquad\qquad\  \cr\  \qquad\qquad\qquad  
 \delta\widetilde{U}_0^*\!\int\!\!{\rm d}t\,{\rm d}\nu\, U_0^2 \exp\left[ -4 \pi i ( \nu\tau + \omega t ) \right]\cr
+\;\;\;\delta\widetilde{U}_0\!\int\!\!{\rm d}t\,{\rm d}\nu\, U_0^*U_0.
\end{eqnarray}
Although the first two terms on the right-hand side of equation 4
are comparable in magnitude, their counterparts in equation 6 are not;
generally the term in $\delta\widetilde{U}_0$ is expected to dominate the term in
$\delta\widetilde{U}_0^*$, because only a small subset of the power in $U_0$
contributes to the coefficient in $\delta\widetilde{U}_0^*$, whereas (by
Parseval's theorem) all of the power in $U_0$ contributes to the
coefficient of $\delta\widetilde{U}_0$. Noting that
$\int\!{\rm d}t\,{\rm d}\nu\, U_0^*U_0 \simeq \int\!{\rm d}t\,{\rm d}\nu\, I(\nu, t)$,
because $|\delta U_0|\ll|U_0|$, we see that we can choose
$\int\!{\rm d}t\,{\rm d}\nu\, U_0^*U_0=1$ at the outset, by appropriately
normalising the dynamic spectrum. Neglecting the conjugate image
(i.e. the term in $\delta\widetilde{U}_0^*$)  then leads to the simple form
\begin{equation}
\delta\widetilde{U}_0=\widetilde{{\cal R}_0 U_0}.
\end{equation}
This result gives us an estimate for the (complex) amplitude of the
scattered wave which is missing from the model, as a function of the
assumed delay and Doppler-shift of that wave. This result has been
derived under the assumption that $\delta U_0$ is dominated by a
single scattered wave, so that the appropriate choices of delay and
Doppler-shift are those values for which the modulus of the right-hand-side
attains its largest value.

The result just derived shows how, given a model electric field
$U_0$, we can improve that model by adding a single scattered
wave component. Moreover the process can clearly be iterated,
so that the restriction embodied in equation 5
-- i.e. that $\delta U_0$ can be approximated by a single plane wave
-- is unimportant: the full spectrum of scattered waves can be
built up iteratively. This is done by adding one new component
to the reference model, $U_0$, on each iteration.
In its overall structure this procedure is similar to that of the
CLEAN algorithm (H\"ogbom 1974) which is commonly employed
in aperture synthesis imaging in radio astronomy; that algorithm
inspired the one presented here.

We need to assume a model as a starting point for an iterative
solution; in the absence of any specific a priori information about
the scattered waves, the most sensible choice is a zero-frequency
($\tau, \omega = 0$) plane wave model. This choice may correspond,
physically, to an unscattered wave. However, we note that the
interference fringe properties depend on delay/Doppler {\it differences,\/}
so our model predicts the same dynamic spectrum for the scattered
wave field $\widetilde{U}_1(\tau, \omega)$ as it does for $\widetilde{U}_2
(\tau, \omega)=\widetilde{U}_1(\tau+\tau_0, \omega+\omega_0)$.
This point can be recognised most easily if we write down one of
the interference terms contributing to $I=U^*U$:
\begin{equation}
\widetilde{u}_k^*\widetilde{u}_j
\exp[2\pi i \{\nu(\tau_j-\tau_k)+(\omega_j-\omega_k)t\}].
\end{equation}
This degeneracy means that there is no information in the dynamic
spectrum on the {\it absolute\/} delay and Doppler shift of the
scattered waves, just as there is no information on absolute phase
($\widetilde{u}_k^*\widetilde{u}_j=|\widetilde{u}_k||\widetilde{u}_j|
\exp[i(\phi_j-\phi_k)]$), and the origin in our model
$\widetilde{U}$ is arbitrary.

\subsection{Weak scattering limit}
To our knowledge only one previous attempt has been made to
derive the scattered wave spectrum from a recorded dynamic spectrum:
B.J.~Rickett (personal communication 2003) derived a solution
$\{\widetilde{u}_j, \tau_j, \omega_j\}$
in the weak scattering limit. In this limit all of the scattered wave components
are of very low amplitude in comparison with the unscattered component,
and to obtain a solution all of the cross-terms $\widetilde{u}_j^*\widetilde{u}_k$
with $j, k\ne0$ are neglected.  This approximation has a simple
correspondence with the approach we have described. Because
the scattered components are all very weak, there is no need to
keep refining the model electric field, $U_0$, and consequently
it is not necessary to iterate the solution: we simply take all of the
field components $\delta\widetilde{U}_0$ returned by equation~7
for the starting model, $U_0=1$. This yields the solution in the weak
scattering limit: $\widetilde{U}=\widetilde{U}_0+\delta\widetilde{U}_0=\widetilde{I}(\tau, \omega)$.
Our approach, by contrast, employs an unscattered reference wave
($U_0=1$) only for the first cycle of the iteration process; all
subsequent cycles differ from the weak scattering limit by including
the identified scattered components in the reference model, and in
general $U_0\ne1$ in equation~7.

\section{An implementation of the algorithm}
To evaluate the procedure described in the previous section we first
tested it on a simple synthetic dynamic spectrum made up of four scattered wave
components plus a small amount of pseudo-random ``noise''.  Whereas
the derivation in \S2 makes no particular assumption about the sampling
of $I(\nu, t)$, our synthetic dynamic spectrum, and the real data discussed
in \S4, were sampled on a regular grid in radio-frequency and time.
In Fourier space, our representation of these data need only employ
wave components with conjugate variables (delay and Doppler-shift)
which lie on a uniformly spaced grid satisfying the Nyquist sampling
criterion; our algorithm employs such a grid.

We found that the algorithm
did indeed locate (in delay-Doppler space) the electric-field
components which we knew to be present, but it did not make accurate
estimates of the (complex) field amplitudes when the components were
first identified. This is not surprising because the procedure described
in \S2 employs simplifying assumptions, so equation 7 is not expected
to give a very accurate estimate of the component amplitudes.
This property would be unimportant
if subsequent iterations refined the values of the component amplitudes
so that the algorithm converged on the correct solution. In fact the
procedure generated spurious components,
instead of converging on the correct values of the actual field components,
thus smearing power around in delay-Doppler space. This behaviour
yields results of low dynamic range and is unacceptable.

Because of the poor performance of the procedure described in
\S2, we decided to refine the algorithm slightly. Instead of simply accepting
the largest Fourier coefficient in equation 7 as the appropriate estimate of the
wave amplitude, we refined the estimate by adjusting the wave amplitude and
phase so as to achieve a least-squares fit of the revised model to the dynamic
spectrum. In this scheme the estimate
provided by equation 7 is used to fix $\tau$ and $\omega$ for the new
component, as before, but the (complex) amplitude indicated by
equation 7 is used only as the starting point for the least-squares
minimisation. By fitting to the data we sidestep many of the concerns
which might otherwise arise in connection with the approximations
made in deriving equation 7 --- linearisation, the single component
approximation, and neglect of the conjugate image. In short the
revised procedure should work well providing only that the largest
Fourier component of ${\cal R}_0U_0$ is also the largest Fourier
component of the difference between the reference field and the
actual field ($\delta U_0=U-U_0$). We found that this algorithm did indeed
perform well, yielding reconstructions which matched the input as closely
as possible (i.e. to within the ``noise''), and did not generate any spurious
components beyond what would be expected given the ``noise'' level
of the ``data'' --- i.e. the algorithm did not limit the dynamic range
of the results. In summary, the successful algorithm follows the procedure
specified in \S2, but at each iteration the amplitude and phase of the
new electric field component (identified via equation 7) is determined
by least squares fitting to the data. The algorithm is described in detail
in the following section.

\subsection{Details of the algorithm}
The algorithm was implemented in IDL, a high-level, commercial
software package which provides straightforward array
manipulation and data display.  The main elements of the algorithm are:

(i) The data are preconditioned; this involves three steps.
First we remove the spectral profile, $B(\nu)$, which is imposed
by the bandpass filter; this is part of our routine data reduction
process, with $B(\nu)$ determined from a calibration data set.
Secondly, the effects of intrinsic fluctuations in the pulsar's flux are
removed, in so far as possible; these fluctuations manifest themselves
as a purely temporal modulation, $f(t)$, with a white noise spectrum.
We dealt with this modulation by forming the frequency-averaged
intensity at each timestep; we then filtered out the high-frequency
components in this quantity, by mutiplying by a Gaussian function
of full-width-half-maximum equal to one half of the Nyquist frequency,
and transformed back to yield a smoothed version of the average
intensity; finally we made our estimate of $f(t)$ from the ratio of these
two quantities. Correcting the data for the effects of pulse-to-pulse
variations is then simply a matter of dividing $I(\nu,t)$ by $f(t)$.
It must be acknowledged that this prescription is not ideal, as it
does not completely eliminate the intrinsic flux variations, and it
also attenuates the wave interference structure slightly. The final step
is to normalise to unit mean intensity. 

(ii) The starting model for $U_0$ is taken to be a wave of unit amplitude
and zero phase, with $\tau, \omega = 0$.

(iii) A new scattered wave component is then identified as
the largest component in equation 7 which has $\tau\ge0$.
Physically we expect that all delays should be non-negative
relative to the unscattered wave; the algorithm can in principle
differentiate between waves of positive and negative delay, so
it should be able to discover this property. However, if the unscattered
wave is very strong, so that $\widetilde{U}\simeq\delta(\tau)\,\delta(\omega)$,
then $\widetilde{U}_0\simeq\widetilde{U}_0^*$ and it is difficult for any
practical scheme to avoid confusion between a scattered wave
and its complex conjugate. As the latter are all equivalent to
scattered waves which have $\tau\le0$, there is potential for
choosing spurious components, and in practice we found that
this did indeed happen. To avoid this problem we imposed the
restriction $\tau\ge0$ on all components.

Our input data consist of measurements of field intensity on a
regular grid in radio-frequency and time -- the dynamic spectrum
-- so for these data the Fourier plane (delay-Doppler) representation
requires only components on a regular, Nyquist-sampled grid.
Consequently we chose our scattered wave components
to lie on a Nyquist-sampled grid in $(\tau, \omega)$.

(iv) The amplitude and phase of the new component are
adjusted by minimising the sum of squares of the difference
between the model and the data. Minimisation of several
component amplitudes/phases is undertaken simultaneously,
using the ``Amoeba'' algorithm described in Numerical Recipes,
which is implemented within the IDL software package.
The `scale' parameter used by the Amoeba algorithm is initially
set to 0.2, when a new component is first introduced, and then
is reduced by a factor 0.7 on each subsequent iteration (i.e.
every time another component wave is added to the model).
Only components whose scale parameters are greater than
some value (we employed $2\times10^{-3}$) are included
in our least-squares minimisation; these criteria mean that
13 scattered wave components are routinely included. In
addition, because the unscattered wave  ($\tau, \omega = 0$)
is often very strong this component is also included in the optimisation,
making 14 wave components in total.

This gradual  ``freeze-out'' of component amplitudes, as the
Amoeba scale parameter gradually decreases, was incorporated
into our algorithm in order that components with similar amplitudes
could be simultaneously adjusted (optimised) in an efficient way.
This aspect of the algorithm has an additional benefit: it helps to
guard against the possibility of spurious parameter determinations
arising from local, rather than global minima found by Amoeba. If
Amoeba finds a local minimum, with correspondingly erroneous
component amplitudes, on a given iteration, it may well find its way
out of the local minimum to reach the global minimum on the
next iteration cycle. Only one of the 14 component amplitudes
ceases to be adjusted on each cycle, so the algorithm has a
fair degree of robustness to the potential problems caused
by local minima in the chi-squared hyper-surface. If, for some
reason, the algorithm were to fix a component amplitude at
a value which is badly in error, then that error may be fixed
in later iteration cycles, as there is no barrier to putting
new components in the same place as existing components
if that is where the largest difference between model and data
occurs.

(v) We used the reduced chi-squared value (chi-squared per
degree of freedom) to measure the success of our model in
fitting the data. Each new scattered wave component that is
added to the model should lower the reduced chi-squared,
if that component is significant; however, once the algorithm
reaches the noise level, the new components are (by definition)
no longer significant, and the addition of any given component
is just as likely to increase the reduced chi-squared as to
decrease it. To reflect this change in behaviour we forced
the algorithm to terminate when 3 successive iterations (i.e.
new scattered wave components) each caused the reduced
chi-squared value of the fit to increase. This criterion causes
the algorithm to terminate very quickly once the noise level is
reached, while remaining robust to the presence of a small
number of insignificant components in the solution (i.e. the
algorithm is not tripped up by isolated insignificant components).
It is possible to check the significance of each wave component
in the solution set $\{\widetilde{u}_j, \tau_j, \omega_j\}$ returned
when the algorithm terminates, and then cull insignificant
components; this would be a reasonable requirement to enforce,
but the fraction of such spurious components is expected to
be very small and to date we have not employed any culling.

\section{Performance with real data}
To be useful, the algorithm must be able to cope with real data.
We tested our code on a dynamic spectrum of the pulsar B0834+06,
taken at an observing frequency of 321.0~MHz with the Arecibo
radio telescope. We used the WAPP backend signal processors to
record 1024 channels across a total of  1.563 MHz of bandwidth.
We sampled the spectrum every 4.096 ms and formed a
pulse-phase-averaged estimate of the on-pulse and off-pulse spectra
which we used to calculate the ON - OFF spectrum.  We then
averaged this over 10~s to form one column in the dynamic spectrum. 
The results of a 45 minute observation are shown in figure 1
in the form of the measured dynamic
spectrum (top left) and secondary spectrum (power spectrum; bottom
left), compared with the corresponding values reproduced by our model
(right). From these results we can see that the model yields a convincing
representation of the data: the deficiencies of the modelling are not
apparent to the eye in either the dynamic spectrum or the secondary
spectrum representations.

\begin{figure}
\includegraphics[width=8.5cm]{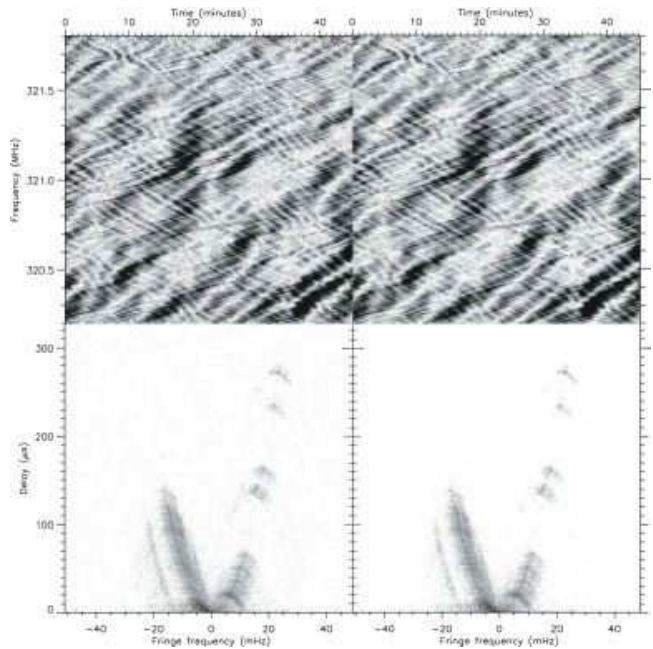}
\caption{Data (left) and model (right) for an observation of PSR~B0834+06,
in a 1.563~MHz band centred on 321.00~MHz. The data were taken with the
Arecibo radio telescope in conjunction with the WAPP backend signal
processing units, on MJD~53009; there are 1024 spectral channels,
and 270 time samples, each of 10 seconds duration. The top panels show
dynamic spectra, while the lower panels show the corresponding secondary
spectra (power spectra of the dynamic spectra). Inverse grey-scale (black
is peak intensity) is used in all cases; the transfer function is linear for
the dynamic spectra, and logarithmic for the secondary spectra. The
signal/noise ratio on each pixel of the observed dynamic spectrum is
2.7, on average, yielding a dynamic range of 63~dB on the corresponding
secondary spectrum. The model reproduces the data well.}
\end{figure}

The model dynamic and secondary spectra shown in figure 1 are
of course derived quantities; the model itself is the set of scattered
waves (electric field components) represented by $\{\widetilde{u}_j,
\tau_j, \omega_j\}$. In figure 2  we show the scattered wave amplitudes,
$|\widetilde{u}(\tau, \omega)|$;
the roughly parabolic relationship between $\tau$ and $\omega$ is
evident in this plot. These results also show how the scattering/refracting
centres are picked out very clearly in this representation of the data
as power concentrations in delay-Doppler space. It is beyond the
scope of this paper to analyse the information present in the
scattered wave solutions, so we will say nothing specific about
the meaning of the results shown in figure 2; nor do we plot the
phases of the various components. Various applications of our
technique are, however, discussed in general terms in \S5.

In fact our scattered wave model for these data is not acceptable,
in the sense that its chi-square value is too large, given the number of
degrees of freedom. There are $1024\times270$ independent pixels
in the dynamic spectrum; in the model there are 8720 scattered waves,
each of which is described by four parameters (amplitude, phase, delay
and Doppler-shift), and the total over all pixels of the squared-residual between
model and data is $2.88\times10^5$, in units of the noise on the
dynamic spectrum. This corresponds to a reduced chi-squared value
($\chi^2_r$, chi-squared per degree of freedom) only slightly larger than unity:
$\chi^2_r=1.19=2.88\times10^5/(1024\times270-8720\times4)$; but a
statistically acceptable model would have a reduced chi-square
value much closer to unity ($|\chi^2_r-1| \la 0.01$), so although
figure 1 is impressive it is possible to do much better. In other
words the differences between model and data are not due
to noise alone.

A clearer test of the performance of the algorithm comes from
differencing the model and observed secondary spectra, as the
latter exhibits a very high dynamic range. We find that this difference
has a dynamic range of 47~dB, compared to the 63~dB dynamic range
of the input data. These figures confirm that the model accounts for the
majority of the structure in the data, but the residuals are far from noise-like.
We expect that substantial improvements in the quality of the fit would be
possible if the 8720 (complex) component amplitudes were simultaneously
optimised. Global least-squares optimisation of our solution has not been
attempted by us. With such a large number of free parameters, a global
optimisation is not an easy task: in a linearised approach the simultaneous
equations we are required to solve have $\sim3\times10^8$ non-zero
coefficients.

\begin{figure}
\includegraphics[width=8.5cm]{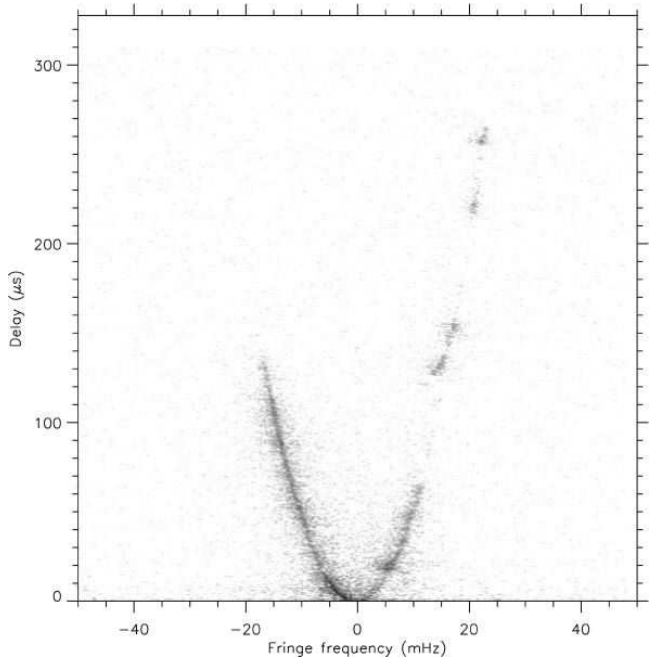}
\caption{The amplitudes, $|\tilde{u}_j|$, of the 8720 scattered
wave components identified by the algorithm described in \S3,
operating on the dynamic spectrum shown in figure 1. These
wave components form the basis from which the model
dynamic and secondary spectra in figure 1 are derived.
The axes are delay ($\tau$), and Doppler shift ($\omega$).
In this figure we can see that the individual scattered waves
cluster tightly around a parabolic locus with $\tau\propto\omega^2$.
An inverse logarithmic grey-scale is used for this figure.
}
\end{figure}

\section{Discussion}
As noted above, the algorithm we have described does not
generate a statistically acceptable description of the data shown in figure 1.
However, it was not obvious at the outset of this study that the
type of spectral decomposition process we sought would work
at all, so the partial success we are able to report is in fact quite
encouraging. We expect that much better fits to data will be
achieved in future with further development of this technique.
In part our confidence stems from the fact that a pulsar dynamic
spectrum can be expected to be accurately modelled by
the form $I(\nu, t)=U^*U$ (once calibration etc.
is taken care of), and the job is simply to find $U$. More
precisely, the job is to find the phase of $U$, since the
amplitude is known directly from $I$. This emphasises the
importance of a fact mentioned in the introduction: this type
of problem, i.e. phase-retrieval, has previously
been addressed in various contexts in the optics literature,
and future work on dynamic spectrum decomposition should
make full use of that resource.  Our
particular application corresponds to the problem of retrieving
a ``complex-valued object'' (in our case, $\widetilde{U}$) from
the modulus of its Fourier transform ($|U|=\sqrt{I}$); this
is recognised as a difficult problem (McBride, O'Leary and Allen 2004).
An acceptable model should yield noise-like residuals (in
both dynamic and secondary spectra), and should satisfy
the constraint that the number of free parameters be
very much less than the number of independent measurements
of the dynamic spectrum. Our current algorithm fails on
the first of these criteria. We expect that a globally optimised
solution -- in which all free parameters are simultaneously
adjusted -- would come much closer to achieving noise-like
residuals, but we have not yet demonstrated this. With such
a large number of free parameters, a global optimisation would
be computationally challenging.

An important limitation of the algorithm we have presented
is that it is restricted to finding scattered waves with $\tau\ge0$.
Under many circumstances this limitation does not cause
problems. However, if there are multiple refracted images
present then problems can arise. To see why, we need to
consider the properties of the starting model: a single component
of unit amplitude at $\tau,  \omega=0$; what does this component
represent? Because the data themselves only carry information
about the {\it relative\/} Doppler-shifts, delays and phases amongst the
various wave components, the choice of origin is arbitrary. In practice,
because the starting model places all of the flux in a single wave,
the origin actually corresponds to the component which contains
the largest flux. If there is only a single refracted image present,
then this component will also be the component with the smallest
delay. However, if there are multiple refracted images present, then
the brightest of these might well not be the path with the smallest
delay. In this circumstance the algorithm will fail to find the
scattered and refracted wave components which have smaller
delays than the starting model -- because they have $\tau<0$
-- and cannot be expected to return an accurate model of
the electric fields, regardless of how many iterations are performed.

Our description of the received signal in terms of scattered wave
components may be used for a variety of purposes; to date we
have recognised three main applications, which we describe in
the following sections.

\subsection{Imaging the ionised ISM}
The scattered wave components are identified by the
values of their delay and Doppler-shift, $\tau$ and $\omega$.
If the scattering occurs in a single, thin screen -- as often seems
to be the case when organised patterns are seen in a dynamic
spectrum (Stinebring et al 2001) -- then there is a direct
relationship between these coordinates and the apparent positions
of the scattering centres. This relationship takes a simple form
in cases where the observed delays are dominated by the
geometric path delay, leading to parabolic features in the
secondary spectrum; this circumstance is quite common
(Stinebring et al 2001; Cordes et al 2005). If the mapping between
$(\tau, \omega)$ and position can be determined, then the scattered
waves can be remapped to give an image of the refracting/scattering
centres. Such a picture would be complex (containing both ampltitude
and phase information), and detailed, and should provide valuable
insights into the ionised component of the interstellar medium.
In particular it will be helpful in elucidating the nature of the
anomalous scattering and refracting screens which are
known to cause a variety of phenomena -- Extreme Scattering
Events (Fiedler et al 1987, 1994); refraction and multiple imaging
in pulsars (e.g. Hewish 1980; Cordes and Wolszczan 1986; Rickett,
Lyne and Gupta 1994; Hill et al 2005);
and intra-day variability in compact radio quasars (Kedziora-Chudczer
et al 1997; Dennett-Thorpe and de~Bruyn 2000; Bignall et al 2003).
These phenomena are poorly understood (Rickett 1991, 2001).

\subsection{Quantifying pulse TOA errors}
If some of the signal arriving at the telescope is delayed,
then a measurement of the pulse Time-Of-Arrival (TOA)
will be affected by this delay. As the scattering
geometry necessarily changes from epoch to epoch,
this is a source of systematic errors in precision pulsar
timing observations. Such studies are important
for tests of fundamental physics -- the equation of state
of matter at nuclear densities, for example -- for low-frequency
gravitational wave detection, and for precision time-keeping
(Foster and Backer 1990).
Eliminating propagation errors in pulse TOAs would therefore
be valuable. By analysing dynamic spectra with the technique
we have presented, we obtain detailed information about the
various paths from source to observer, allowing us to quantify
the effects of multi-path propagation on the pulse TOAs.
In turn this means that we should be able to eliminate this
contribution to the systematic errors in TOAs if high resolution
dynamic spectra are recorded in tandem with the timing information.

We can envisage two different schemes for correcting the
scattering errors in pulse TOAs. The simplest scheme would
be to compute the net delay due to all the known paths, and
then subtract this delay from the corresponding TOA. The
second scheme is much more ambitious: if the pulse TOA
is determined from off-line reduction of baseband data, then
we can, in principle,  process those data in such a way
that the electric fields from the various scattered paths are
coherently recombined with the unscattered signal. This
generalises and extends the concept of coherent
de-dispersion (Hankins and Rickett 1975), whereby the
``filtering'' imposed by propagating the signal through a
cold plasma is precisely removed by applying the inverse
filter. Coherent de-dispersion is now routinely used to process
baseband data on radio pulsars. By analogy with a
phase-conjugate mirror, which eliminates wavefront
errors by exactly reversing light propagation paths,
we can term a coherent recombination of the scattered
signal ``Virtual Phase Conjugation'' (VPC). VPC has the
potential to completely remove the signal filtering imposed
by (multi-path) interstellar propagation, and thus to eliminate
the associated contributions to pulse TOA errors. By the
same token, VPC should permit studies of pulsar microstructure
(e.g. Hankins 1996) at very high time-resolution.

\subsection{High resolution imaging of the source}
In \S5.1 we noted that the scattered wave decomposition
yields, fairly directly, an image of the scattering medium.
With some further development it should also permit
interferometric imaging of the pulsar itself.
To see why, we need only recall how terrestrial radio
interferometers operate: the correlation between pairs
of signals is evaluated as a function of baseline, i.e.
separation between the antennas. A high visibility
amplitude for a given source means that the coherent
patch is larger than the baseline length, so the
source must be small. Conversely, if the visibility
amplitude is small then the baseline is longer than
the size of the coherent patch, and we have resolved
the source on this baseline. Exactly the same considerations
apply to the interference between the scattered waves
discussed here; in our case the interferometric baseline
is simply the transverse separation of the paths at the
location of the scattering screen.

In the particular approach we have described in this paper the
scattered waves are identified under the assumption of
a point-like source (see \S2) with an infinitely large
coherent patch, so that all pairs of scattered waves, $j,k$
yield fringes of amplitude $|\widetilde{u}_j||\widetilde{u}_k|$.
For very large scattering angles, this approximation must
break down, and in this case the data can be used to image
the source in a manner directly analogous to terrestrial
radio astronomy, namely by quantifying the fringe visibility
as a function of baseline length. This technique is fundamentally
similar to previous investigations which used
interstellar scattering to constrain the size of pulsar radio-emission
regions (Gwinn 2001; Gwinn et al 1997; Wolszczan and Cordes 1987).
There are two main advantages of the method proposed here.
First, it makes use of a lot more information --- all the contributing
propagation paths are elucidated, thus permitting many more
constraints on the model brightness distribution of the pulsar.
Secondly, by knowing the geometry of all the contributing paths
we can re-order the visibility data into the usual coordinate
system (the ``$u,v$ plane'') used for radio astronomical imaging
with terrestrial interferometers. From that position the many
powerful concepts, techniques and tools developed for synthesis
imaging can be brought to bear on the problem of imaging
via interstellar scattering.

\subsection{Interstellar holography}
B.J.~Rickett (personal communication, 2003) has previously
noted the close analogy between his scattered wave solution,
derived in the weak scattering limit (\S2.1), and Gabor holography.
The scattered wave decomposition described in this paper goes
beyond the weak scattering approximation, but the relationship
to holography remains strong.  In both cases the dynamic spectrum
can be regarded as an in-line (Gabor) hologram of the interstellar medium.
In the weak scattering limit the hologram is recorded with a plane
reference wave, in effect -- the strong, unscattered wave -- and
can be reconstructed with such a wave to
yield an image of the scattering medium. In this paper we have
considered the more general circumstance where the amplitudes
of the scattered components are not negligibly small, and their
mutual interference must be taken into account. In this case the
dynamic spectrum is analogous to a hologram recorded with an
aberrated reference wave, and reconstructing an image is no
longer so straightforward. Our approach will permit images to
be reconstructed computationally for this situation. The
connection with Gabor holography is worth bearing in mind
because many powerful techniques have been developed in
that domain, and some may be useful in interstellar holography.

\section{Conclusions}
We have considered the possibility of representing pulsar
dynamic spectra in terms of an identifiable collection of electric
field components whose mutual interference yields the observed
intensity structure. Such a representation is of interest
because of the direct link between the properties of
the field components and the characteristics of the contributing
propagation paths through the interstellar medium. An algorithm
for achieving this decomposition has been derived, implemented,
and tested on high-quality data; it works well, although the resulting
model is not formally an acceptable representation of the data as
the residuals are not noise-like.
Further development of this technique should permit
insights into a variety of issues relating to interstellar wave
propagation and the physics of pulsars.

\section{Acknowledgments}
We have benefitted from helpful discussions with several
colleagues: Simon Johnston, Barney Rickett, Jim Cordes, 
Bill Coles, Curtis Asplund, Ben Stappers, Willem van Straten
and Leon Koopmans. At the University of Sydney this work
was supported by the Australian Reseach Council, at the
Kapteyn Institute and ASTRON by the Netherlands Organisation
for Scientific Research, and at Oberlin by the National Science Foundation.

\end{document}